\documentclass[fleqn,usenatbib]{mnras}

\usepackage{newtxtext,newtxmath}
\usepackage[T1]{fontenc}
\usepackage{ae,aecompl}
\usepackage{graphicx}

\usepackage{amsmath}
\usepackage{lipsum}

\usepackage{multirow}
\usepackage{dcolumn}% Align table columns on decimal point

\title[\textit{Lazarus Stars}: Star-lifting as a life extension strategy]{\textit{Lazarus Stars}: Numerical investigations of stellar evolution with star-lifting as a life extension strategy}

\author[Scoggins \& Kipping]{Matthew~T.~Scoggins$^{1}$\thanks{E-mail: mts2188@columbia.edu}
David Kipping,$^{1}$
\\
$^{1}$Department of Astronomy, Columbia University, New York, NY, 10027\\}

\date{Accepted XXX. Received YYY; in original form ZZZ}

\pubyear{2022}

\begin{document}
\label{firstpage}
\pagerange{\pageref{firstpage}--\pageref{lastpage}}
\maketitle

\begin{abstract}
    The aging and gradual brightening of the Sun will challenge Earth's habitability in the next few billion years. If life exists elsewhere in the Universe, the aging of its host star similarly poses an existential threat.  One solution, which we dub a Lazarus star, is for an advanced civilization to remove (or star-lift) mass from their host star at a rate that offsets the increase in luminosity, keeping the flux on the habitable planet(s) constant and extending the lifetime of their star. While this idea has existed since 1985 when it was first proposed by Criswell, numerical investigations of star-lifting have been lacking. Here, we use the stellar evolution code MESA to find mass {\it vs.} age and $\dot{M}$ {\it vs.} age relations which would hold the flux on surrounding planets constant. We explore initial mass ranging from $0.2\,{\rm M}_{\odot}$ to $1.2\,{\rm M}_{\odot}$. For most stars with a mass initially below about $ 0.4 {\rm M}_{\odot}$, we find that star-lifting increases their main-sequence lifetimes up to $500$\,Gyr until they approach the hydrogen burning limit and star-lifting is no longer possible. For more massive stars, star-lifting increase main-sequence lifetimes by 1\,Gyr to 100\,Gyr, though they still enter the red-giant phase. For example, a Sun-like star has a main-sequence lifetime which can be increased by up to 3\,Gyr. This requires a mass-loss rate of about $0.05\, {\rm M}_{\mathrm{Ceres}}$ per year. We compare star-lifting to other survival strategies and briefly discuss methods for detecting these engineered stars.
\end{abstract}

\section{Introduction}
The upper limit for the window of a planet's habitability is set by the lifetime of its host star. If the increasing luminosity during the red-giant phase isn't enough to compromise the biosphere of the planet, the death of the host star surely spells the end for any living beings in its planetary system. A sufficiently intelligent civilization has several pathways for survival. Shorter-term solutions that cope with the changing luminosity of their host star include terraforming their planet, migrating to another planet in their system, or something like planetary migration, such as asteroid deflection which increases the orbital radius and keeps the flux on the planet constant \citep{Korycansky_2001}. A longer-term solution could involve star-lifting (SL), the process of removing mass from the host star in order to slow down the rate of nuclear fusion and prevent a dramatic increase in luminosity. This idea was first popularized in 1985 by David Criswell \citep{Criswell_1985}, was further developed by \citet{Beech_2008}, then a simple SL analysis was explored for a Sun-like star by \citet{Matloff_2017}. Matloff assumes the mechanism for SL would involve a laser which increases the star's mass ejection via stellar winds and explores several scenarios, with the most conservative approach resulting in about a 3 percent mass loss over a 600\,Myr period.

%If this removed mass stays interior to the orbit of the habitable planet, the orbital radius of the planet should remain approximately unchanged so keeping the star's luminosity constant ({\it isoluminosity}) is equivalent to keeping the flux on the planet constant. If the mass is ejected out of the system, the planet will drift and mass should be removed at a slower rate ({\it isoirradiance}) in order to keep the flux on the planet constant.

In this paper, we set aside the engineering complications of SL and derive the required mass {\it vs.} age and $\dot{M}$ {\it vs.} age relations which maintain a constant bolometric irradiance on a habitable planet. This paper is organized as follows. In section \ref{sec:methods} we describe two methods for star-lifting ({\it isoluminosity} and {\it isoirradiance}), introduce MESA and share our numerical techniques for implementing SL in MESA. In section \ref{sec:results} we present our results for the engineered mass {\it vs.} age and $\dot{M}$ {\it vs.} age relations, give a worked example for a Sun-like star and compare extended lifetimes to the natural lifetimes for each SL method. In section \ref{sec:discussion} we discuss these results, compare SL to other habitability-extending pathways and consider detection methods.

\begin{figure*}
 \includegraphics[width=\columnwidth]{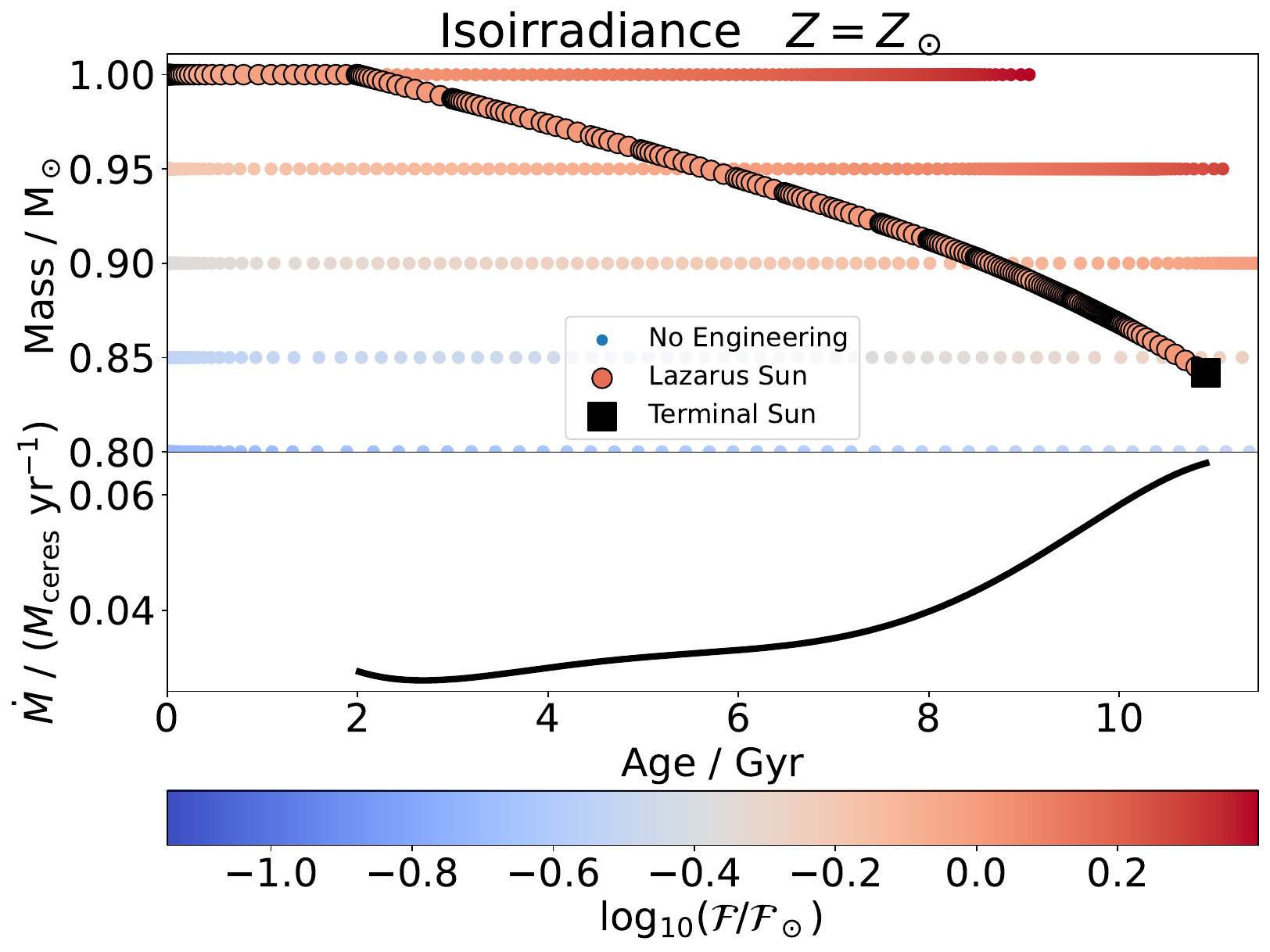}
 \includegraphics[width=\columnwidth]{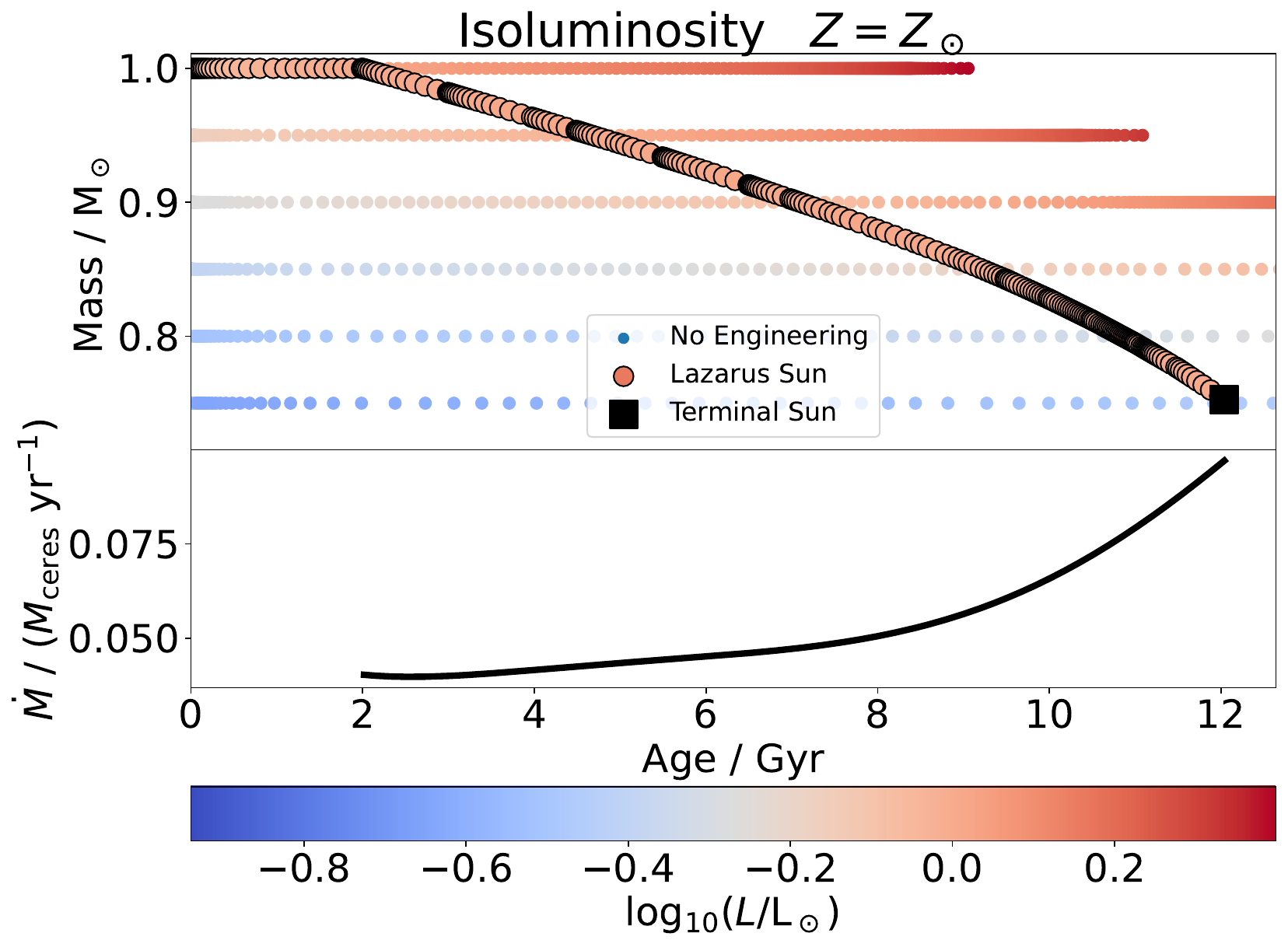}
 \caption{An example of the mass {\it vs.} age relation that a star like our Sun would follow if SL began in the past, $A_{\rm start}=2.0$\,Gyr, for {\it isoirradiance} (left) and {\it isoluminosity} (right). The natural evolutionary tracks are plotted behind this, with a nearly constant mass (nearly horizontal evolution) but increasing $\mathcal{F}$ or $L_*$ (coloured) as the stars age. For a given track, we interpolate both mass and age for a fixed $\mathcal{F}$ or $L_*$. Our Sun's main-sequence lifetime is roughly 9\,Gyr (at least until the central hydrogen fraction reaches $10^{-10}$), while a Lazarus Sun would stay on the MS for $11$\,Gyr with {\it isoirradiance} and $12$\,Gyr with {\it isoluminosity}. The required $\dot{M}t$ is shown in the bottom panels, in units of\,$M_{\rm ceres}$yr$^{-1}$. Terminal Sun marks the points where our Lazarus Suns reach a central hydrogen fraction below $10^{-10}$}
 \label{fig:worked_example}
\end{figure*}

\section{Methods}
\label{sec:methods}

\subsection{Accounting for changes in orbital radius during SL}
SL could be implemented in several ways. We divide these into two classes. If SL is implemented in a way that relocates mass from a star on to compact objects in a close orbit with that star, keeping the mass interior to the planet and spherically symmetric, the orbit of the planet would not change by Newton's shell theorem \citep{Newton_1687}. In this case, keeping flux on the planet constant is equivalent to keeping the luminosity of the host star constant and we call this implementation {\it isoluminosity}.

Alternatively, SL could be implemented in a way that dramatically increases the rate the star loses mass through stellar winds \citep{Matloff_2017}.  A mechanism like this would result in roughly continuous mass loss which is gradual enough to keep the habitable planet in a stable circular orbit, though this would cause the orbital radius of the planet to increase as the mass leaves the volume enclosed by the planet's orbital radius. Assuming this is done so that there is net torque on the habitable planet, the planet's orbital angular momentum will be conserved. 

For planet mass $m_{\rm p}$, orbital radius $r$, stellar mass $M_*$ and gravitational constant $G$, angular momentum of the planet follows $J_{\rm p} = m_{\rm p}rv_{\rm p} = m_{\rm p} \sqrt{GM_*r}$ meaning the quantity $M_*r$ is kept constant. With $M_* \propto \frac{1}{r}$, the incident flux on the planet $F_{\rm inc}$ follows
\begin{align}
    F_{\rm inc} \propto \frac{L_*}{r^2} \propto L_*M_*^2
\end{align} 
where $L_*$ is the luminosity of the host star. So, for a given age, we solve for the mass such that $\mathcal{F} \equiv L_*M_*^2$ is held constant. Though both implementations hold bolometric irradiance constant for a planet at any orbital radius, we call this implementation where the lifted mass is moved outside the orbital radius {\it isoirradiance}. 

Both methods have their merits (see \S~\ref{sec:discussion} for a  comparison of methods for extending habitability) so it will be difficult to predict which mechanism an advanced civilization would prefer. Our work explores both mechanisms for mass loss, where {\it isoluminosity} sets an approximate upper bound for the required mass loss rate $\dot{M}$ and {\it isoirradiance} sets an approximate lower bound.

\subsection{MESA stellar evolution code}
We calculate mass {\it vs.} age relations using the Modules for Experiments in Stellar Astrophysics code (MESA) \citep{Paxton_2010, Paxton_2013, Paxton_2015}. We explore stars with initial mass $M_0$ such that $0.2 \leq M_0/{\rm M}_\odot \leq 1.2$ with increments of $\delta M_0{=}0.05\,{\rm M}_\odot$ and metallicity $Z \ {\epsilon} \  \{0.01, 0.1, 1\}$ in units of $Z_\odot = 0.0122$. Exploring this range for both implementations results in a total of 126 engineered evolutionary tracks.

We star-lift using the \texttt{mass\_change} variable in MESA, where this variable sets the rate of accretion per year, but allows negative values which account for mass loss. To implement this, we allow stars to follow a natural evolution until $A_{\rm start} = 2$\,Gyr, after which point $L_*$ or $\mathcal{F}$ is held constant. In order to hold $L_*$ or $\mathcal{F}$ constant, we start with the model for the star at $2$\,Gyr, fix a timestep $\delta t$, and evolve that model several times until time $2$\,Gyr + $\delta t$ with each evolution holding different values for \texttt{mass\_change}. We then interpolate across the evolved values of $L_*$ (or $\mathcal{F}$) at our desired constant value to solve for the required \texttt{mass\_change}, where a final evolution for this timestep is carried out with this value. We continue to make these timesteps until the star reaches the end of its life, where the whole process is repeated for smaller $\delta t$ until this final evolution converges. We stop the evolution when the star approaches the end of the main sequence (MS) and the central hydrogen fraction drops below $10^{-10}$ or when maintaining {\it isoluminosity} or {\it isoirradiance} is no longer possible. This second condition occurs when a star approaches the hydrogen burning limit and begins to cool and dim without losing any mass.

\section{Results}
\label{sec:results}
\subsection{Mass {\it vs.} age relations}

Fig.~\ref{fig:worked_example} shows a worked example for a Sun-like star, applying both {\it isoirradiance} and {\it isoluminosity}, with SL beginning at $A_{\rm start}=2.0$\,Gyr. The derived mass {\it vs.} age relation is plot against the natural evolutionary tracks which are coloured according to $L_*$ (or $\mathcal{F}$). The natural evolutionary tracks are nearly horizontal, because the total mass lost during the main sequence is much smaller than the initial mass. Every timestep after $A_{\rm start}$ is fit with a 6th-order polynomial, which is then used to calculate $\dot{M}$ in the bottom panels. {\it Isoirradiance} requires a slightly more modest star lifting rate of $\dot{M} \approx 0.03\,\ M_{\rm Ceres} \text{yr}^{-1}$, and extends the Sun's MS lifetime to about $11$\,Gyr, increasing the total MS lifetime by 2\,Gyr. {\it Isoluminosity} initially requires a larger SL rate, resulting in a MS lifetime of about $12$\,Gyr, increasing the total MS lifetime by 3\,Gyr. We mark the point when our modified Lazarus Sun reaches a central hydrogen fraction below $10^{-10}$ and star lifting is no longer viable beyond this point. We call this point Terminal Sun.

\begin{figure*}
 \includegraphics[width=2\columnwidth]{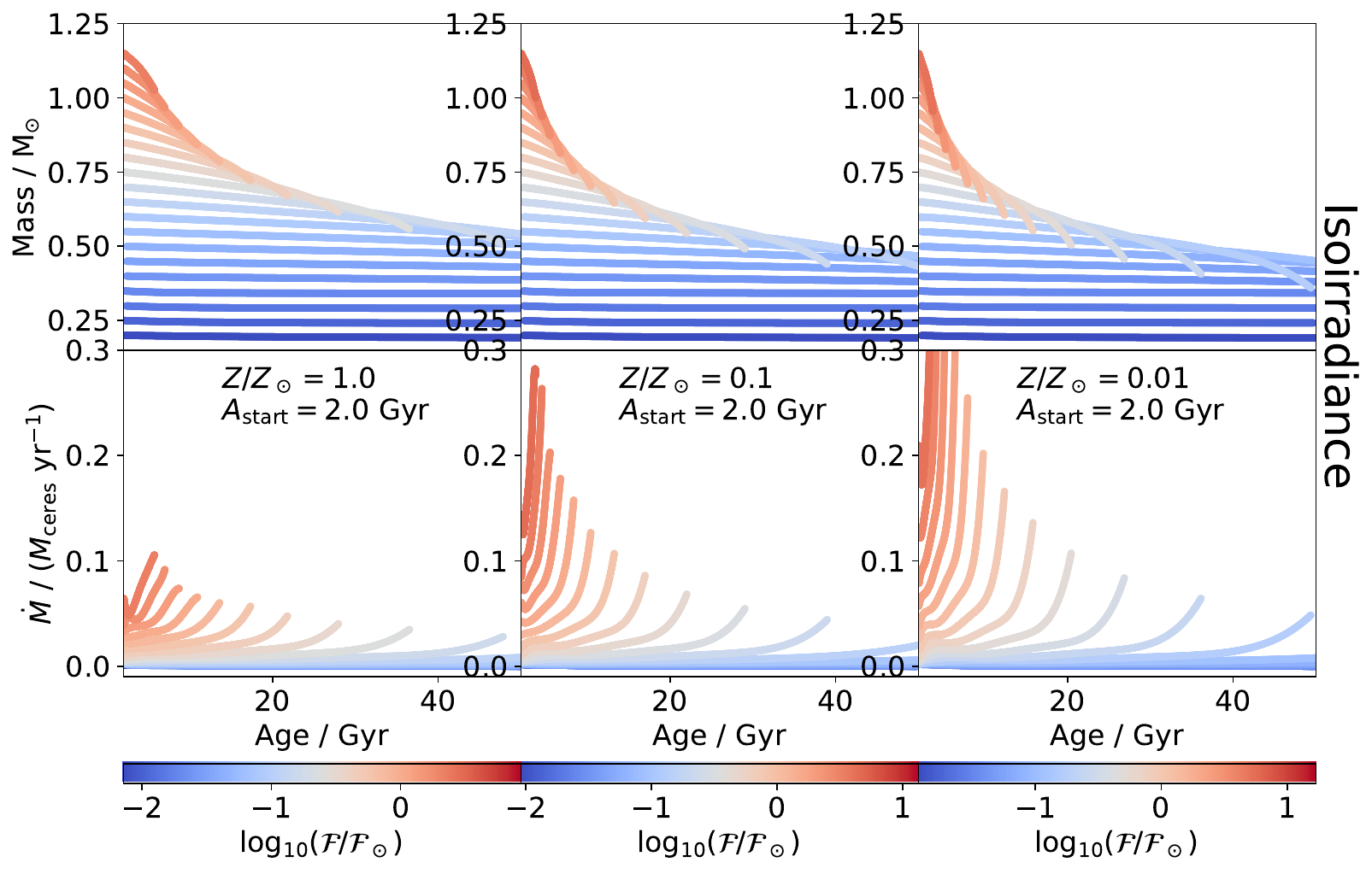}
 \includegraphics[width=2\columnwidth]{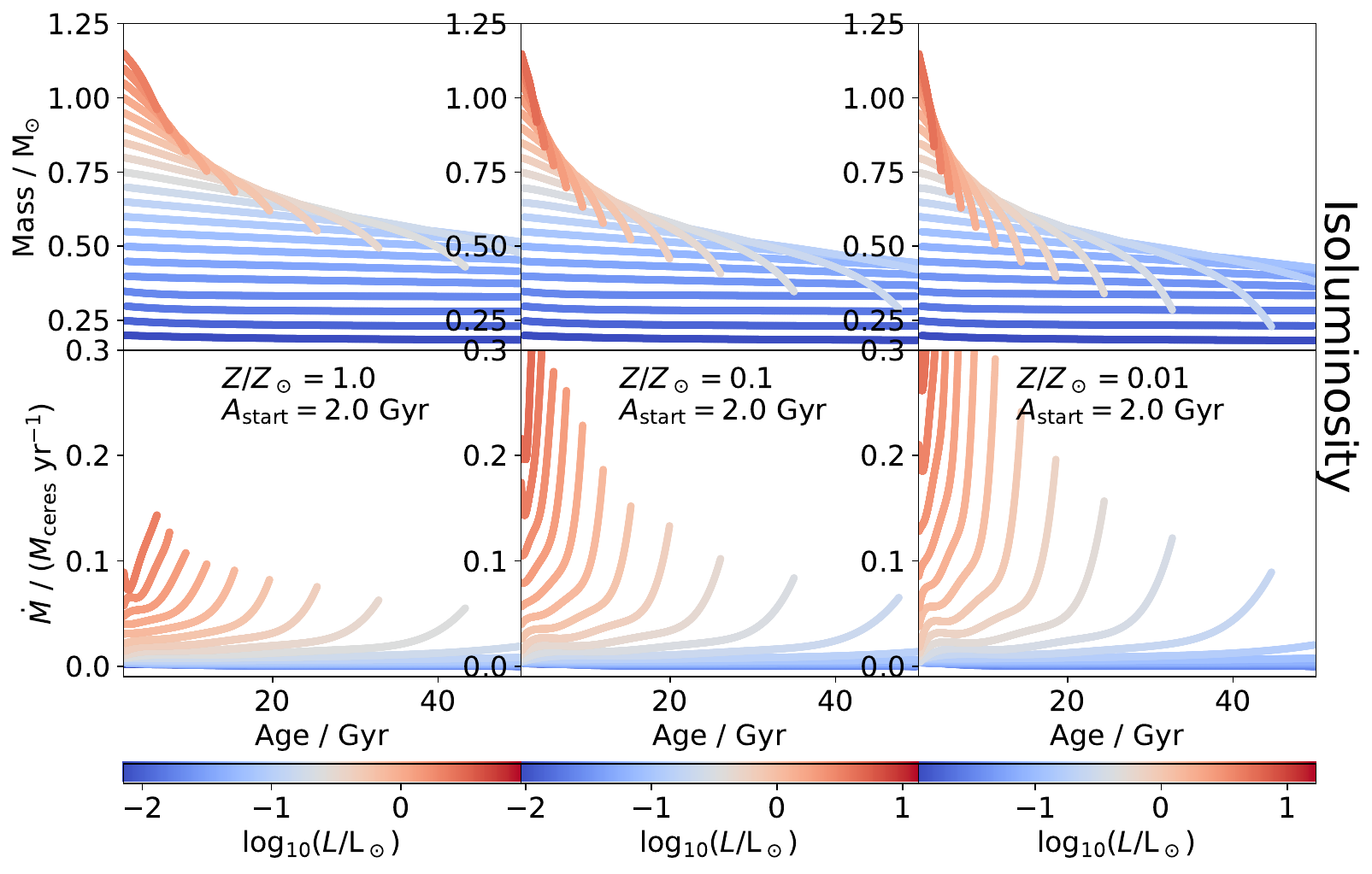}

 \caption{Mass {\it vs.} age and $\dot{M}$ {\it vs.} age for {\it isoirradiance} (top) and {\it isoluminosity} (bottom) with $Z\epsilon \{0.01, 0.1, 1\}Z_\odot$ with $Z_\odot = 0.0122$. Colours indicate ${\rm log}_{10}(\mathcal{F}/\mathcal{F}_\odot)$ ({\it isoirradiance}) and ${\rm log}_{10}(L/{\rm L}_\odot)$ ({\it isoluminosity}) when SL begins, $A_{\rm start} = 2$\,Gyr, and held constant across the engineered evolution. Mass {\it vs.} age lines are determined by holding $\mathcal{F}$ or $L_*$ constant during MESA evolution. $\dot{M}$ is calculated with the derivative of a 6th-degree polynomial fit to mass  {\it vs.} age and shown in units of $M_{\rm Ceres}$ yr$^{-1}$, where $M_{\rm Ceres} = 9.1\times 10^{20}\,$kg.}
 \label{fig:m_vs_a}
\end{figure*}

Derived mass {\it vs.} age and $\dot{M}$ {\it vs.} age relations across all initial masses explored are shown in Fig.~\ref{fig:m_vs_a} with $Z/Z_\odot \epsilon \ \{0.01,0.1, 1\}$ and $A_{\rm start}=2$\,Gyr. We focus on the first 50\,Gyr because most stars with initial mass greater than $0.8 {\rm M}_\odot$ do not live beyond this window. We find that the initial $\dot{M}$ required increases with initial mass and, for a fixed track, $\dot{M}$ typically increases with age. We choose a small $A_{\rm start}$ in order to capture the behaviour of these higher mass stars, which would see very little benefit to a later start time (because most would be nearing the end of their MS life) and low-mass stars live long enough that choosing a later $A_{\rm start}$ has a negligible effect on their engineered evolution.

\subsection{Extended lifetimes}

\begin{figure*}
 \includegraphics[width=1.95\columnwidth]{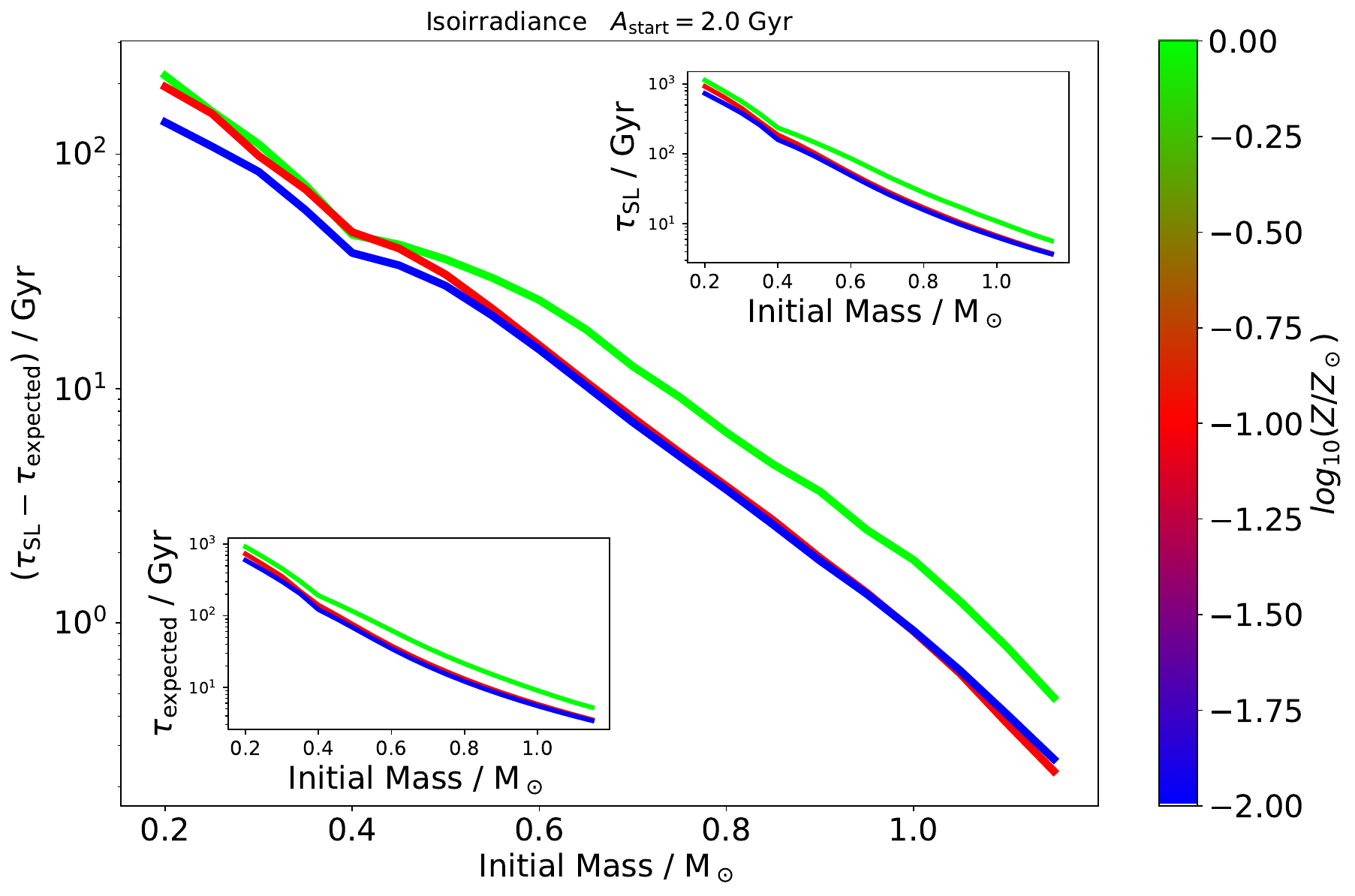}
 \includegraphics[width=1.95\columnwidth]{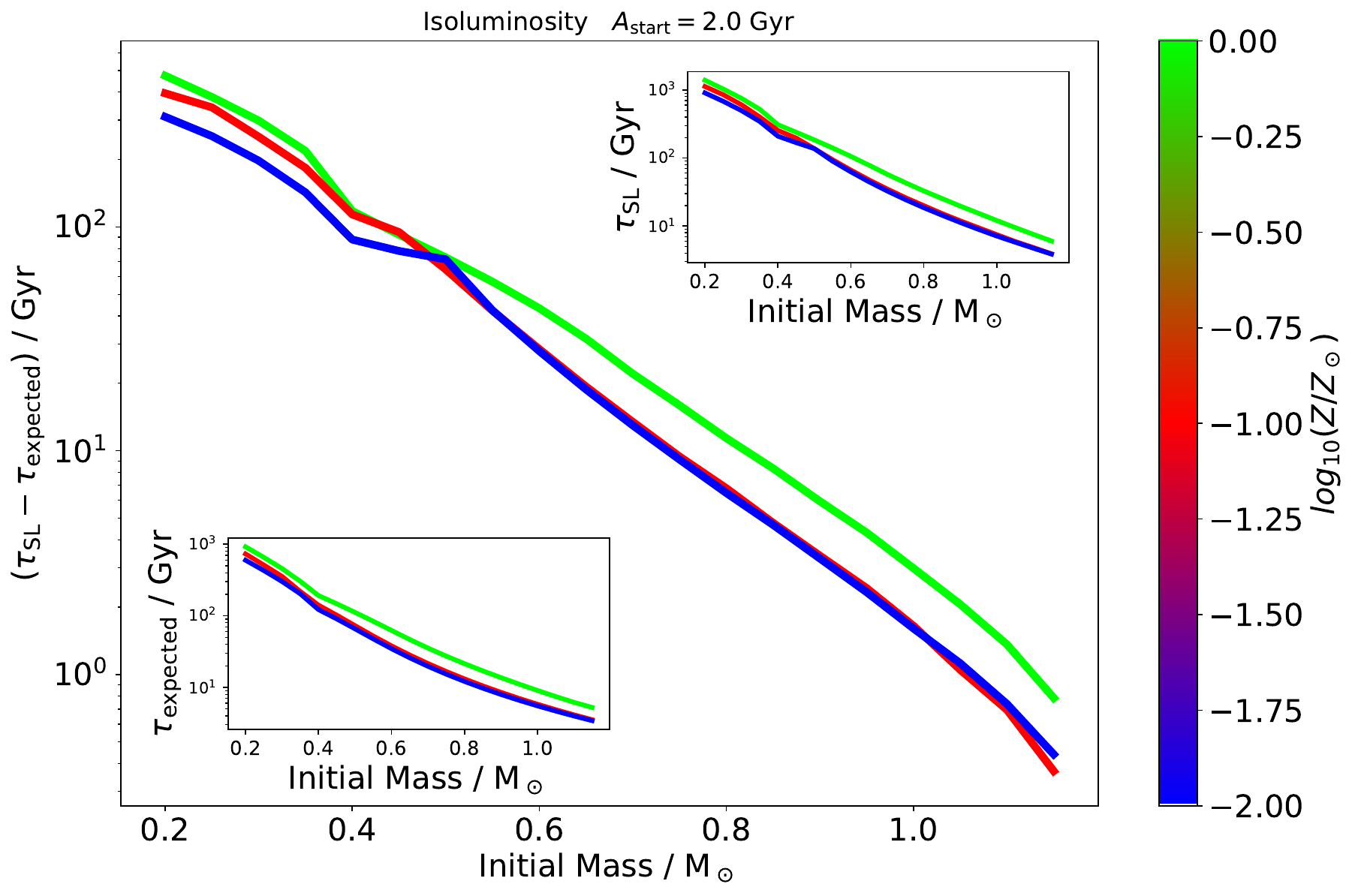}
 \caption{Initial mass {\it vs.} {\large $\tau_{\rm SL} -\tau_{\rm expected}$} for the main-sequence lifetime with SL ($\tau_{\rm SF}$) and without SL ($\tau_{\rm expected}$). Colour indicates metallicity. Top: {\it Isoirradiance}, holding $\mathcal{F} = L_*M_*^2$ constant. Bottom: {\it Isoluminosity}, holding $L_*$ constant.}
 \label{fig:lifetime_ratios}
\end{figure*}

In order to show the benefit of SL for all stars in the explored range, we compare MS stellar lifetimes in Fig.~\ref{fig:lifetime_ratios} for $A_{\rm start}=2$\,Gyr, plotting the difference between the star-lifted MS lifetime $\tau_{\rm SL}$ and the expected MS lifetime $\tau_{\rm expected}$ against the initial mass of the star. We also subplot $\tau_{\rm SL}$ and $\tau_{\rm expected}$ directly. 

Both methods see similar patterns, though they differ in magnitude. For a fixed initial mass, {\it isoluminsoity} results in a greater increase in MS lifetime, because this method removes more mass than {\it isoirradiance} and is more efficient at slowing fusion. For both methods, higher-mass stars with $M_* \gtrsim 0.8 {\rm M}_\odot$ see a modest lifetime increase of $0.1$\,Gyr to $10$\,Gyr. As the initial mass grows, the increased lifespan decreases, eventually approaching $0$ as $A_{\rm start}$ becomes greater than the main-sequence lifetime. Lower mass stars see an increase of $10$\,Gyr to $500$\,Gyr. Generally, the lifetime gains scale with the natural MS lifetime of the star.

\begin{figure*}
 \includegraphics[width=1.5\columnwidth]{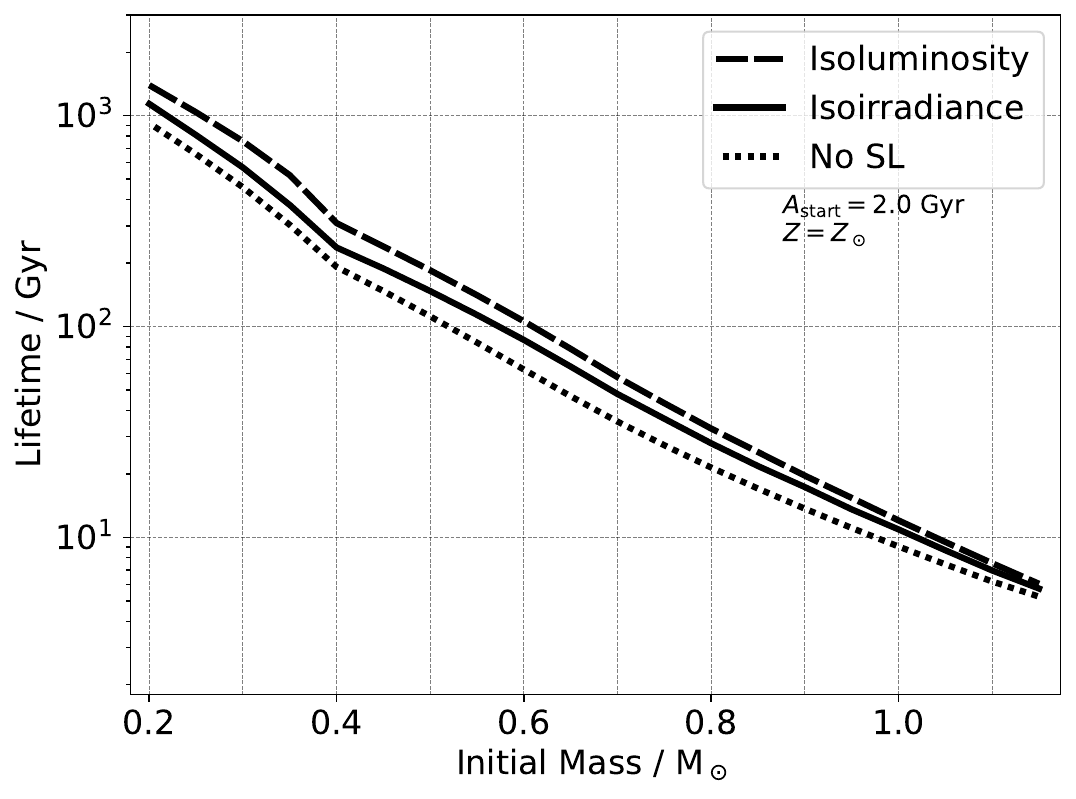}
  \caption{Initial mass {\it vs.} lifetime for both methods (dashed for {\it isoluminosity} and solid for {\it isoirradiance}) and natural evolution without SL (dotted), where $Z=Z_\odot$ and $A_{\rm start} = 2$\,Gyr.}
 \label{fig:terminal_age}
\end{figure*}

Fig.~\ref{fig:terminal_age} shows the lifetimes of stars with $Z=Z_\odot$ as a function of initial mass for both methods, compared to the lifetime without SL. We see convergence at the higher end of the mass range as $A_{\rm start}$ approaches the main-sequence lifetime of these stars. We also find convergence as initial mass approaches the hydrogen-burning limit, where maintaining {\it isoirradiance} or {\it isoluminosity} is only possible for a brief time and provides little benefit. We also show the movement of a few of these stars in mass-luminosity space in Fig.~\ref{fig:funny_HR}.

\section{Discussion}
\label{sec:discussion}

\subsection{Star-lifting {\it vs.} Alternative Methods}

Before considering detection of SL, it is important to compare the benefit and energy requirements of SL to other methods, with the intention of gauging the likelihood that SL is chosen over these other methods. While it is difficult to predict what an advanced civilization would be capable of, and what an implementation of alternative life-extension strategies might look like, we can make rough approximations for their energy requirements.

The minimum power required to SL at a rate of $\dot{M}$, moving the material to a radius $a$, can be approximated with the difference in potential energy of the final and initial location of the material
\begin{align}
    \dot{E} &\approx (U_f - U_i)/dt = GM_* \dot{M} [-\frac{1}{a}+ \frac{1}{R_*}] \approx GM_* \dot{M} \frac{1}{R_*},
\end{align} where we have assumed $a \gg R_*$ for {\it isoluminosity} (or $a\rightarrow \inf$ for {\it isoirradiance}). For our Sun, with a required SL rate of $\dot{M}  \approx 5{\times}10^{19}$kg $\text{yr}^{-1}$, this would require $\dot{E} \approx 10^{38}$ erg $\text{yr}^{-1}$, or about $10^{-4} {\rm L}_\odot$. We emphasize that this is a lower bound for the power requirement which assumes perfect efficiency. In a more detailed calculation, \citealt{Matloff_2017} finds the power required to lift mass from the Sun's photosphere to be $1.9 \times 10^{20}$ erg $\text{kg}^{-1}$. This results in $\dot{E}$ about $10^{40}$ erg ${\rm yr}^{-1}$, or about $0.01 {\rm L}_\odot$.

Taking place over billions of years, the total energy required would be on the order of $10^{47}$ erg, at least. This is several orders of magnitude larger than the total energy required for migrating Earth via asteroid deflection in order to keep the incident flux constant, about $10^{40}$ erg \citep{Korycansky_2001}. Alternatively, migration could be achieved with a large reflective sail which is much less massive than the required asteroid \citep{McInnes_2002}. Though SL is much more expensive, the energy required for this is readily available from the star and offers a long-term solution by extending the lifetime of the host star, whereas the upper limit for habitability via orbital migration is still set by the natural MS lifetime.  Further, assuming more than one habitable planet in the system through luck or terraforming, SL extends the lifetime of the whole planetary system by keeping the flux on every planet in that system constant. For a multi-planetary civilization SL may be preferred because a technique similar to orbital migration requires a transfer of orbital energy from one planet to another. This makes it difficult to preserve multiple habitable planets in the same system. Orbital migration could also lead to a destabilization of asteroids or orbits of other planets \citep{Korycansky_2001}. Spin up or spin down torque considerations also have to be made with the orbital migration technique. {\it Isoirradiance} may face similar difficulties, which may be an argument for a civilization to prefer {\it isoluminosity}.

Rather than SL, the civilization might choose to relocate. Relocating to a planetary system in the local Milky Way (e.g. 100 ly away) with peak speed $0.1c$ would require more than $10^4$ yr. A journey like this for a human-like civilization with $10^{10}$ inhabitants, assuming relocation of each inhabitant and a ship mass of $5\times 10^4$ kg per inhabitant (the ISS mass-capacity ratio, though this is likely very conservative for an interstellar journey) would require a ship of mass about $10^{15}$ kg, if not much larger. Ignoring the Solar System escape velocity ($ v_{\rm esc} \ll 0.1c$), the total energy required for migration of the entire civilization is roughly $KE \approx \frac{1}{2}10^{15} (3\times 10^7)^2 = 4.5 \times 10^{36}$ erg, though this is likely extremely conservative. While this is cheaper than SL, the logistics of achieving near light-speed travel, finding a fuel source for accelerating this much mass, and survival of the inhabitants during the journey may make this infeasible for a large population. Migrating a small fraction of the population may be preferred, while the majority of the population left behind would benefit from SL.

Terraforming is another option. Among other requirements, generating a magnetic field would be an extremely energy demanding necessity. \citet{Bamford_2022} find the minimum energy stored in a magnetic field to create a habitable Mars-like planet to be $10^{24}\,$erg, and this doesn't include the energy required to ramp up the magnetic field. They find that kick-starting Mars' iron core into an active magnetic dynamo would require $10^{33}$ erg. While technically possible, this is ultimately another short term solution where the upper limit for habitability is set by the MS lifetime of the host star.

Another life extension strategy involves dumping metal-rich objects on to a star, increasing metallicity and slowing fusion, assuming the metals could be mixed with the core (restricting this application to fully convective stars). This could be achieved by kicking objects near the host star out of their orbit and should be relatively energy efficient, though the total amount of material in the planetary system may be a limiting factor. For our own Solar System, we could dump at most a fraction of a percent of a Solar mass on to the Sun. Though this would increase the host star lifetime by an amount roughly equal to the natural lifetime discrepancy between the original and enriched star, the luminosity of the star would still increase with time and a relocation method like asteroid deflection or a reflective sail would be required.

\subsection{Maximizing stellar lifetimes}
Though there are several methods to extending the habitability of a planetary system, we can combine them to get the most out of the host star. We find that both star-lifting methods lead to stars which approach the hydrogen burning limit, though it is possible to SL at much higher rates in order to approach this limit for more massive stars. The limit for the most massive star that could be pushed into the hydrogen burning limit is set by the maximum star-lifting rate, which is determined by the energy available. For the initial mass range explored in this work, the stars radiate much more energy than required for {\it isoirradiance} or {\it isoluminosity} so even a Sun-like star could be engineered into a red dwarf with a more aggressive SL rate. This would cause a decrease in luminosity, though inward orbital migration could offset this.

\subsection{Detection of star-lifting}

\begin{figure*}
 \includegraphics[width=1.5\columnwidth]{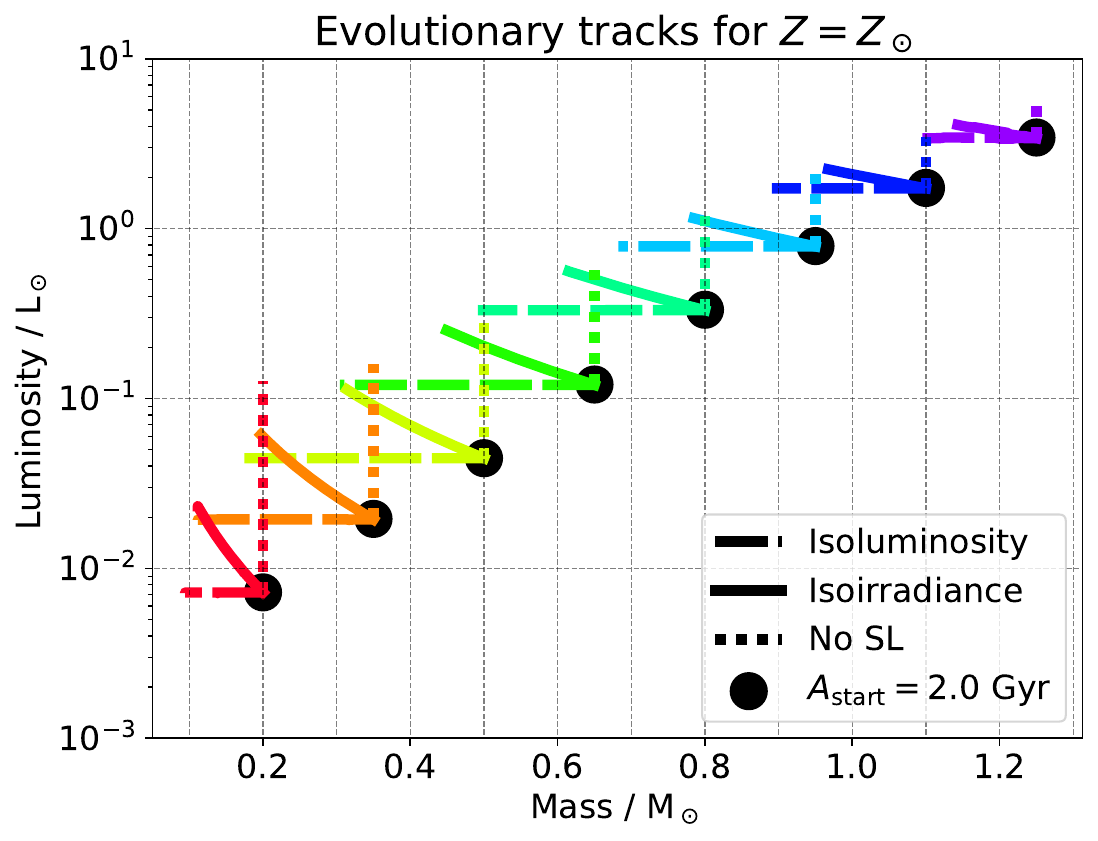}
  \caption{Evolution in Mass--Luminosity space for both methods of SL (dashed for {\it isoluminosity} and solid for {\it isoirradiance}) and the natural evolution without SL (dotted). {\it Isoluminosity} holds luminosity constant by definition, so movement is perfectly horizontal and we see this method result in the smallest final MS masses. {\it Isoirradiance} allows for a slight increase in luminosity in order to offset the increased orbital radius of the host planet after mass is ejected out of the planetary system. Natural evolution appears vertical, though there is a slight mass loss during the MS due to natural stellar winds.}
 \label{fig:funny_HR}
\end{figure*}

Though difficult, there are several ways which we could observe SL. Implementing {\it isoluminosity} would likely cause dips in the apparent brightness as the star-lifted material orbits between the distant observer and the host star. This was considered by others (e.g. \citealt{Boyajian_2016, Wright_2015}). An {\it isoirradiance} implementation could also cause noticeable effects. If implemented through an increased stellar wind, it may be possible to detect the unusually high wind directly. Some of the most massive stars explored in this work require a mass-loss rate which matches or exceeds previous lower limits for stellar wind detection, about $ 10^{-10} {\rm M}_{\odot}\text{yr}^{-1}$ \citep{Lamers_1999}. Further, extremely long term observations could detect these unusual changes in mass, where a star's mass and luminosity evolution might follow one of the unusual tracks shown in Fig.~\ref{fig:funny_HR}. 

SL could also take the form of an unusually large population of low-metallicity stars in extremely old clusters -- stars which should have expired given the age of the cluster may still exist if they have been engineered. Though these stars would be hard to distinguish from a star which naturally has a similar mass, age and metallicity, a population distribution which is skewed towards the lower-mass range or which is completely missing red giants could hint at SL, assuming the civilization has engineered on the Galactic scale.

Contradictory ages derived from gyrochronology and asteroseismology could also be a sign of SL. For example, if we correctly calculate the age via asteroseismology for an engineered star, we won't necessarily find a similar age via gyrochronology. The spin of the star for a fixed mass and age could vary wildly depending on the implementation of SL and where that removed mass is placed. Inferring different ages in a binary pair may also hint at SL.

\section*{Acknowledgements}
We thank the reviewer, C Tout, for helpful comments and feedback. We thank Jason Wright, Ben Cassese, M.-M. Mac Low, Jamie Tayar, and Doug Scoggins for helpful comments and discussion. We thank the authors of the freely available code Matplotlib \citep{Hunter:2007}, SciPy \citep{2020SciPy-NMeth} and NumPy \citep{harris2020array}.

Special thanks to donors to the Cool Worlds Lab, Mark Sloan,
Douglas Daughaday,
Andrew Jones,
Elena West,
Tristan Zajonc,
Chuck Wolfred,
Lasse Skov,
Graeme Benson,
Alex de Vaal,
Mark Elliott,
Methven Forbes,
Stephen Lee,
Zachary Danielson,
Chad Souter,
Marcus Gillette,
Tina Jeffcoat,
Jason Rockett,
Scott Hannum,
Tom Donkin,
Andrew Schoen,
Jacob Black,
Reza Ramezankhani,
Steven Marks,
Gary Canterbury,
Nicholas Gebben,
Joseph Alexander,
Mike Hedlund,
Dhruv Bansal,
Jonathan Sturm,
Rand Corporation,
Leigh Deacon,
Ryan Provost,
Brynjolfur Sigurjonsson,
Benjamin Paul Walford,
Nicholas De Haan,
Joseph Gillmer,
Emerson Garland,
Alexander Leishman,
The Queen Road Foundation Inc,
Brandon Thomas Pearson,
Christian Slutz,
Scott Thayer,
Marjorie Waters,
and Joseph Gillmer.

\section*{Data Availability}
The code developed for star lifting and data produced during the preparation of this manuscript is available at this \href{https://github.com/mscoggs/star_lifting}{github repository}.

\bibliographystyle{mnras}
\bibliography{references} 
\bsp
\label{lastpage}
\end{document}